# Domain nucleation across the metal-insulator transition of self-strained V$_2$O$_3$ films

Alexandre Pofelski[1†], Sergio Valencia[2†], Yoav Kalcheim[3], Pavel Salev[4], Alberto Rivera[5], Chubin Huang[3], Mohamad A. Mawass[2], Florian Kronast[2], Ivan K. Schuller[6], Yimei Zhu[1]* and Javier del Valle[7]*

[1]Condensed Matter Physics and Material Science Department, Brookhaven National Laboratory, Upton, NY 11973, USA.
[2]Department of Spin and Topology in Quantum Materials, Helmholtz-Zentrum Berlin für Materialen und Energie, 12489 Berlin, Germany
[3]Department of Material Science and Engineering, Technion - Israel Institute of Technology, Haifa 32000, Israel
[4]Department of Physics and Astronomy, University of Denver, Denver, CO 80210, USA.
[5]GFMC, Departamento de Física de Materiales, Facultad de Física, Universidad Complutense, Madrid, 28040 Spain
[6]Department of Physics and Center for Advanced Nanoscience, University of California San Diego, La Jolla, CA 92093, USA.
[7]Department of Physics, University of Oviedo, C/ Federico García Lorca 18, 33007 Oviedo, Spain

[†]These authors contributed equally to this work

*Corresponding authors: zhu@bnl.gov; javier.delvalle@uniovi.es

**Abstract**

Bulk V$_2$O$_3$ features concomitant metal-insulator (MIT) and structural (SPT) phase transitions at T$_C$ ~ 160 K. In thin films, where the substrate clamping can impose geometrical restrictions on the SPT, the epitaxial relation between the V$_2$O$_3$ film and substrate can have a profound effect on the MIT. Here we present a detailed characterization of domain nucleation and growth across the MIT in (001)-oriented V$_2$O$_3$ films grown on sapphire. By combining scanning electron transmission microscopy (STEM) and photoelectron emission microscopy (PEEM), we imaged the MIT with planar and vertical resolution. We observed that upon cooling, insulating domains nucleate at the top of the film, where strain is lowest, and expand downwards and laterally. This growth is arrested at a critical thickness of 50 nm from the substrate interface, leaving a persistent bottom metallic layer. As a result, the MIT cannot take place in the interior of films below this critical thickness. However, PEEM measurements revealed that insulating domains can still form on a very thin superficial layer at the top interface. Our results demonstrate the intricate spatial complexity of the MIT in clamped V$_2$O$_3$, especially the strain-induced large variations along the c-axis. Engineering the thickness-dependent MIT can provide an unconventional way to build out-of-plane geometry devices by using the persistent bottom metal layer as a native electrode.

**Introduction**

Metal-insulator transitions (MITs) have recently experienced renewed interest. The large changes in electrical, optical, and structural/mechanical properties associated with the MIT have made them attractive for applications in novel technologies such as passive cooling [1], neuromorphic computing [2–8], probabilistic computing [9,10], optoelectronics [11], and micromechanical actuators [12]. In addition, the unresolved debate about the role played by the electronic and lattice degrees of freedom across the MIT keeps drawing the attention of a large community in condensed matter physics [13–22].

Among the many materials featuring a MIT, $V_2O_3$ is considered a paradigmatic example. Un-doped, un-strained $V_2O_3$ features a MIT at around 160 K, across which the resistivity changes over six orders of magnitude [23,24]. The MIT is concomitant with a structural phase transition (SPT) between the high-temperature corundum and low-temperature monoclinic structures (Fig. 1a) [25]. The corundum metallic phase has a hexagonal unit cell. Across the SPT, the corundum $c$ lattice parameter shortens (-0.36%), the $c$-plane area expands (+1.29%), and the $c$ axis tilts slightly, i.e., the angle between the $a$ and $c$ axes becomes less than 90º. This monoclinic tilt can happen along three equivalent directions, and therefore three monoclinic twins are possible.

In thin films, $V_2O_3$ can be strained and confined laterally by the substrate. This puts constraints on the SPT, which can no longer take place freely. Because of the SPT, the nucleation of insulating/metallic domains within the film results in compressive strain being applied to the neighboring areas, which can either promote or hinder the MIT in those areas [26]. When insulating domains nucleate, they usually arrange in an alternating pattern that minimizes strain [8,19,20]. Domain self-organization due to strain accommodation is not unique to $V_2O_3$ and is often observed in multiple systems featuring an SPT, such as $VO_2$ [27], rare earth nickelates [28,29], manganites [30], and FeRh [31].

The case of $V_2O_3$ is especially striking given the big change in lattice constants, which is largest on the base of the hexagonal unit cell [25]. In 001-oriented $V_2O_3$ films, that base is parallel to the plane, and is therefore confined in size by the substrate. As a result, the film cannot expand, and the MIT is strongly suppressed. The observed degree of suppression can range from an incomplete film transformation across the MIT to a full suppression of the insulating state and persisting metallic phase down to 0 K.

The intense self-strain that develops during domain nucleation provides an unconventional platform to study the pressure-temperature phase diagram, with access to both positive and negative pressure regimes, and can stabilize novel phases without any external driving force [26,32]. For instance, it was recently shown that the paramagnetic insulator phase in pure $V_2O_3$, not stable at room pressure, can be stabilized in thin films by imposing strain on the (100) corundum plane [32]. Furthermore, epitaxy-induced structural locking has been proposed to decouple the MIT from the SPT. Several works have reported such decoupling in related compound $VO_2$ [13,33] and, more recently, also in (001)-oriented $V_2O_3$ [34,35].

Direct visualization of how domains nucleate and grow is key for understanding and exploiting the complex phase diagram of these highly correlated systems. Planar (top view) imaging of the transition has been thoroughly reported using different probes [20,36]. These techniques, however, only offer a partial description since they are not sensitive to variations along the film thickness. A complete picture requires cross-sectional imaging that shows the phase transition extent across the thickness of the film.

In this work we provide a complete picture of domain nucleation and growth in strongly self-strained, (001)-oriented $V_2O_3$ films. We combined two techniques that allowed us to capture planar as well as transversal snapshots of the MIT: photoelectron emission microscopy (PEEM) and scanning electron transmission microscopy (STEM). We observed that insulating domains first nucleate at the surface of the film where strain is lowest and grow downwards and laterally until they stop at around a critical thickness of 50 nm from the substrate/film interface. The existence of a critical thickness prevents thinner films from nucleating any insulating phase withing their interior. PEEM imaging, however, revealed a thin superficial layer on the film's surface where the MIT can take place, even for the thinnest films. Domain formation here shows very different features compared to the bulk of the film.

Our results offer a comprehensive spatially resolved picture of the MIT in strongly strained systems. This information is fundamental for proper interpretation of macroscopic measurements of the MIT properties, and for the strain and epitaxy control and stabilization of novel electronic and structural phases. On the practical level, engineering the thickness-dependent MIT can simplify the fabrication of out-of-plane switching devices, as the persistent metallic phase region below the critical thickness can be potentially utilized as a native bottom electrode, alleviating the need for conducting substrates or conducting buffer layers.

**Growth, transport, and X-ray diffraction**

$V_2O_3$ films were epitaxially grown on top of c-cut (001) sapphire ($Al_2O_3$) substrates using RF magnetron sputtering from a stoichiometric $V_2O_3$ target. Growth was done in an 8 mTorr Ar atmosphere with a substrate temperature of 700 °C. After growth, the samples were thermally quenched at a rate of around 90 °C/min. Four samples of different thicknesses were grown: 300 nm, 100 nm, 50 nm and 18 nm. Film thickness was confirmed by X-ray reflectivity measurements (Fig. 1b). Specular X-ray diffraction (XRD, Fig. 1c) shows that $V_2O_3$ films are compressively strained by the underlying substrate, the compression being larger for the thinner films, as expected.

Figure 1d shows resistance vs temperature for all four samples, while Figure 1e shows resistance normalized to the 300 K resistance for each sample, so that transport properties can be better compared. In all cases, the usually observed five-to-seven orders of magnitude $V_2O_3$ MIT is strongly suppressed. The degree of suppression is dependent on the film thickness, as reported before [26]: thinner films show a smaller upturn in resistance, indicating a lower insulating phase fraction in the low-temperature range.

The impact of film thickness on the monoclinic insulator phase formation is further confirmed by low temperature XRD measurements. Figure 2 and supplementary Figure S1 show reciprocal space maps in the vicinity of the (119) and (006) Bragg peaks, respectively. Two samples are shown: 300 nm and 50 nm. For each case, multiple temperatures between 300 K and 93 K are plotted. For the thicker film, the single corundum peak loses intensity and partially splits into several peaks as the temperature is lowered. These peaks correspond to the different monoclinic twins of the insulating phase. Therefore, as the temperature goes down, part of the film transitions from corundum to monoclinic. Notice however, that the central corundum peak does not disappear: the MIT transformation is incomplete. This explains the moderate resistance upturn in Figures 1d and 1e below the $T_C \sim 160$ K.

A very different scenario takes place for the 50 nm sample. The single corundum peak remains at all temperatures, and no sign of monoclinic phase can be detected in the reciprocal space maps. The corundum peak does, however, lose some intensity at lower temperatures, as can be better seen in the θ-2θ scans of supplementary Figure S2. This implies that at least some fraction of the corundum phase disappears as the temperature is lowered, even though it does not seem to add to the monoclinic phase fraction.

**STEM measurements**

To understand the spatial distribution of domain nucleation, we performed Cryo-STEM measurements. Thin films were sliced into 120 nm thick lamellas by means of a Focused Ion Beam and were subsequently imaged using the LN2 Mel-Build sample holder in a double aberration-corrected JEOL ARM 200F cold FEG microscope, operating at 200 keV in STEM mode (more details in the supplementary).

Figure 3 shows Low Angle Annular Dark Field (LAADF) images of the 300 nm $V_2O_3$ film from 300 K down to 99 K. This imaging mode is especially sensitive to local lattice distortions such as those surrounding defects and grain boundaries. Columnar vertical film growth is readily observed at all temperatures, giving weak contrast variations. Monoclinic areas give a much stronger contrast due to the large lattice orientation difference, and start being visible at 165 K and below.

Nucleation starts from the top and expands downwards and laterally. The monoclinic domain expansion slows down as the temperature approaches 100 K. Importantly, monoclinic domains never reach the bottom substrate-film interface and stop their growth around 50 nm from it, which sets the critical thickness of the domain propagation depth inside the film. Supplementary Figure S3 shows similar measurements for the 50 nm $V_2O_3$ film, i.e., the film right at the critical thickness. No monoclinic phase contrast is visible for any temperature in the STEM imaging, which is in good agreement with XRD measurements in Fig. 2b.

To further confirm the nucleation of monoclinic domains in the 300-nm-thick $V_2O_3$ film and to verify the absence of monoclinic regions in the interior of the 50-nm-thick film, we used a 4D-STEM method, which allowed us to obtain local diffraction patterns with 5 nm spatial resolution. Figure 4a shows one of such patterns taken in a monoclinic domain of the 300 nm thick film. Note that in addition to the array of intense peaks, similar to the one observed for the metallic corundum structure, weaker satellite peaks are discernible in between. These are a hallmark of the monoclinic distortions. Figure 4b shows a local diffraction pattern of the 50 nm film, at 125 K. Extra monoclinic peaks are completely missing, unambiguously ruling out the presence of the monoclinic phase. By repeating this diffraction acquisition in every spot in the field of view it is possible to construct the phase maps shown in Figure 4c, which plot the intensity of the satellite peaks circled in Fig. 4a. From these maps, we can conclude that no monoclinic phase is observed at any temperature in the interior of the 50-nm-thick $V_2O_3$ film.

**Strain analysis**

To understand why the monoclinic domains do not propagate below the 50 nm critical distance from the film-substrate interface, we used the 4D-STEM data to perform strain analysis (more information in the supplementary). The obtained strain maps for the 300 nm film at 293 K are shown in Figure 5a. $E_{xx}$ refers to strain along the horizontal (in plane) direction, while $E_{zz}$ refers to the vertical (out of plane) direction.

It is important to note that strain values are calculated with respect to the lattice parameters of the sapphire substrate, not with respect to relaxed $V_2O_3$. Both metallic $V_2O_3$ and sapphire have corundum structure, but $V_2O_3$ has a larger lattice parameter: 4.1% / 7.8% larger in-plane / out-of-plane, respectively, according to the literature's bulk crystal structure parameters [37,38]. Strain values in Figure 5 are calculated with respect to the substrate. This means that a measured $E_{xx}$ below 4.1% implies compressive strain, since the lattice parameter would be below that of bulk $V_2O_3$, while $E_{xx}$ above 4.1 % indicates tensile strain. Our calculated $E_{xx}$ is around 3.5% near the substrate interface, so at that location $V_2O_3$ is subject to a strong in-plane compression.

A clear vertical dependence for $E_{xx}$ can be identified: compressive strain is strongest near the film-substrate interface and slowly decreases along the z direction. This implies that $V_2O_3$ is compressed along the in-plane direction due to the sapphire substrate clamping, but the film relaxes with increasing thickness, probably due to the proliferation of defects. As expected from the Poisson effect, $E_{zz}$ follows the opposite behavior, being higher near the substrate and decreasing towards the top of the film. These trends are present at different temperatures (Figs. 5a and 5b) and can be better appreciated in the vertical line profiles of Figure 5c.

$E_{xx}$ on the film's top surface is around 4.4%, implying that, far from being compressed, $V_2O_3$ at that location is actually subjected to tensile strain. While this might seem counterintuitive due to the smaller lattice parameter of sapphire, it can be explained by considering the different thermal expansion coefficients of both materials [25,38].

Film growth is done at high temperatures (700º C), where lattice parameters are significantly larger than at the measurement temperature. The $V_2O_3$/sapphire lattice mismatch is even higher at 700º C than at room temperature, so the film grows very compressively strained. The top part of the film, however, will be less compressed that the interface. As the sample is cooled down after growth both materials contract, but $V_2O_3$ contracts faster [25,38], therefore relieving some of the compression. This decompression can get to the point that, for thick films, strain on the top surface changes sign due to the expanding film underneath, resulting in tensile forces [32]. Tensile in-plane strain at the top interface also leaves an imprint on $E_{zz}$ due to the Poisson effect. $E_{zz}$ falls below the bulk value of 7.8%, meaning the c-axis lattice parameter is shorter than bulk on the top of the film. This can be also seen in the XRD measurements on Figure 1c: the 006 $V_2O_3$ peak of the 300 nm film is shifted to higher angles than that of bulk $V_2O_3$.

For the insulating phase to nucleate, the film needs to accommodate a large in plane expansion associated with the corundum-to-monoclinic transition. Such strain accommodation is easier at the top interface, where compressive forces are lowest (or even tensile). Figure 5b shows $E_{xx}$ and $E_{zz}$ at 125 K obtained at the same location as the map in Fig. 5a. Two monoclinic domains can be clearly identified, leaving an imprint on strain. $E_{xx}$ is much higher and $E_{zz}$ much lower within the domain. From these results we can interpret that insulating areas are subjected to a strong compressive stress, which is larger the closer to the substrate, forcing domains to acquire a wedge shape. Stress increases fast below the critical thickness, preventing the insulating domains from ever reaching the bottom substrate-film interface. While here we obtained a critical thickness of 50 nm, its specific value is expected to depend on the substrate choice and its lattice mismatch with the film. Lower mismatches will likely give rise to lower critical thicknesses.

While STEM data shows no monoclinic phase in the interior of the 50 nm $V_2O_3$ film, resistance data in Figure 1b shows an upturn in resistivity at low temperatures, and θ-2θ scans in supplementary Figure S2 show a slight intensity decrease of the corundum structural peak. This suggests that there might be a small fraction of the film that can turn insulating. A closer inspection of the corresponding STEM data (Fig. S3) shows an extremely narrow dark contrast $V_2O_3$ layer on the film surface, right beneath the gold protective layer. A similar layer is observed for the 300 nm film (Fig. 3). Due to its small thickness (< 3 nm) and proximity to the gold film, we are not able to gather much information about this layer using STEM.

**PEEM measurements**

X-ray PEEM, being a surface sensitive technique with limited probing depth (2-3 nm), is an ideal tool to study the properties of the upper surface layer of $V_2O_3$. Shinning x-rays with photon energies corresponding to the resonant Vanadium L-edge induces strong photoemission from the sample surface. The number of electrons ejected is proportional to the x-ray absorption (XAS). Since different $V_2O_3$ phases feature different XAS spectra, photoelectron intensity maps can reveal the distribution of insulating/metallic domains [36]. To investigate the possible transition in the surface layer of the 50-nm-thick $V_2O_3$ film, we performed temperature-dependent PEEM measurements at the UE49-PGMa beamline at the BESSY II synchrotron.

Metallic $V_2O_3$ has two *d* electrons and three overlapping bands at the fermi surface: two degenerate $e_g$ bands arising from *d* orbitals oriented in the plane, and a $a_{1g}$ band coming from a *d* orbital pointing out of plane [39]. As a result, there is some x-ray linear dichroism (XLD): the XAS spectrum differs slightly depending on the orientation of the electric field vector of the incoming linearly polarized beam. Figures 6a and 6c show XAS at 190 K for the 300 nm and 50 nm films, respectively. XAS were measured using radiation polarized either horizontally (H) or vertically (V), which in our experimental setup corresponds to the electric field vector being almost orthogonal and almost parallel to the sample´s surface, respectively. We say that it is 'almost' perpendicular or parallel because the incoming beam has a grazing incidence angle of 16°.

Within the metallic state, XLD = V/H is close to 1. There is little dichroism. When $V_2O_3$ turns monoclinic at low temperature, the $a_{1g}$ band is pushed up, away from the fermi surface, losing occupancy. Consequently, the system becomes more anisotropic and the XLD is no longer close to 1. This can be seen in Figures 6b and 6d, which are taken at 44 K. From them, we can see that XLD is largest around 519 eV. Unlike in XRD and STEM measurements, XLD shows large changes as function of temperature, not only for the 300 nm sample but also for the 50 nm one.

Figure 7a shows space resolved XLD images of the 50 nm thick film, for different temperatures, for a photon energy of 519.2 eV. While the images are featureless at high temperatures, domains can be clearly distinguished at lower temperatures. Interestingly, these domains appear continuously, becoming increasingly more intense, but showing little changes in size. This is observed both in warm up (Fig. 7a) and cool down (Fig. S4). The zoom into a specific domain shown in Fig. S5 allows to see this effect better.

Three XLD intensity levels can be appreciated, corresponding to three different types of domains. To better understand their nature, we divided the image into three different regions of interest, according to their intensity level. XAS spectra were calculated by integrating over each of these regions. Figures 7b and 7c show the XAS taken with in-plane (V) and out-of-plane (H) polarized light. The XAS of the three domains is similar for light polarized out of plane, but clearly different for in-plane polarization. Hence, it is reasonable to interpret the presence of these domains as originating from the three possible insulating, monoclinic twins. Twins have the same out of plane projection, and therefore similar XAS for out-of-plane polarization. But their tilts into different in-plane directions have different projections with respect to the incoming in-plane polarized beam, yielding different XAS for in-plane polarization.

**Discussion**

Our combined PEEM + STEM measurements suggest, on 001-oriented $V_2O_3$ films, the presence of a thin surface layer where monoclinic domains can form. Superficial domain nucleation seems to happen in a rather different fashion, as compared to bulk nucleation. There are very limited changes in domain size as temperature is decreased far below $T_C$ (Fig. S5) and, contrary to bulk domains, their contrast emerges in a continuous way, a feature more reminiscent of second-order transitions. While we do not fully understand why this layer forms, we anticipate it could be caused by large strain relaxation, or a potential lattice reconstruction near the film's surface. The presence of this surface layer might explain the resistance upturn at low temperatures, and the small intensity decrease of the corundum structural peaks in the XRD measurements.

These results show that there is a very strong vertical dependance on the nucleation of insulating domains across the MIT. Not taking this into account could lead to incorrect interpretations. For instance, our XRD results show that 50 nm $V_2O_3$ films do not feature any monoclinic phase (albeit with a small reduction in corundum peak intensity). But our PEEM measurements, which are sensitive to the electronic structure, suggest the formation of an insulating phase, or at least a phase with different orbital symmetry. Without considering variations along the vertical direction, one might have been erroneously led to identify a decoupling between the MIT and the SPT.

From the practical point of view, our observation of a critical thickness can enable an unconventional way to fabricate out-of-plane MIT switching devices. Presently, most studies of the MIT switching utilize a planar device geometry. Planar geometry works well for exploring individual device performance, but it is not optimal for large scale integration. Fabrication of out-of-plane devices requires growing the MIT film on a conducting substrate or buffering the insulating substrate with a conducting layer, which is often detrimental to the MIT film growth as structurally matching conducting substrates or buffer layers are not readily available. Our findings that under appropriate strain conditions, only the top film region undergoes the MIT, while the bottom region always remains metallic alleviate the need of having a conducting substrate/buffer layer. An out-of-plane switching device can be, in principle, made by fabricating a top metal electrode and utilizing the persistent metallic bottom region as a native electrode. As many MIT materials undergo a coinciding SPT, it can be expected that critical thickness of the transition propagation into the interior of the film can be engineered in other materials beyond $V_2O_3$ by appropriately choosing the film-substrate epitaxial relation to impose constraints on the SPT and force the bottom region of the film to remain in the metallic state.

**Conclusions**

We have performed a thorough characterization of domain nucleation and growth across the phase transition in self-strained, (001)-oriented $V_2O_3$ films using a combination of STEM and PEEM measurements. We observed that domain nucleation starts where strain is lowest, at the surface of the film. As the temperature is lowered, insulating monoclinic domains grow downwards and laterally, without ever reaching the bottom part of the film likely due to the high compressive stress. For films less than 50 nm thick, insulating domains cannot nucleate in the interior of the film. However, PEEM measurements reveal the existence of a superficial layer at the top electrode in which insulating phase can form, regardless of film thickness. Surprisingly, surface domain formation shows very different features when compared to that in the interior of the film. Surface domains do not experience large changes in size with varying temperature, and their contrast emerges gradually, resembling a continuous phase transition. These results give fundamental spatially resolved insight into the MIT of strongly self-strained systems. In turn, this is crucial information for stabilizing high pressure phases in these systems, and for the development of MIT-based devices, key for several novel technologies.

**Acknowledgements**

Sample growth and STEM measurements were funded as part of the Quantum Materials for Energy Efficient Neuromorphic Computing (Q-MEEN-C) Energy Frontier Research Center (EFRC), funded by the U.S. Department of Energy, Office of Science, Basic Energy Sciences under Award # DE-SC0019273. The electron microscopy work done at BNL was also supported by the US-DOE, Basic Energy Sciences, Materials Science and Engineering Division under contract No. DE-SC0012704. We thank the BESSY II synchrotron for allocation of beamtime with proposal number 201-09373-ST. Lab-based XRD measurements were supported by the European Union's Horizon Europe research and innovation program under grant agreement No. 2031928 – "Highly Energy-Efficient Resistive Switching in Defect- and Strain- Engineered Mott Insulators for Neuromorphic Computing Applications". Views and opinions expressed are however those of the authors only and do not necessarily reflect those of the European union or the European research council. J.d.V. was supported by the Spanish Ministry of Science through a Ramón y Cajal Fellowship (Grant No. RYC2021-030952-I) and by the Asturias FICYT under Grant No. AYUD/2021/51185 with the support of FEDER funds.

# References

[1] K. Tang, K. Dong, J. Li, M.P. Gordon, F.G. Reichertz, H. Kim, Y. Rho, Q. Wang, C.-Y. Lin, C. P. Grigoropoulos, A. Javey, J.J. Urban, J. Yao, R. Levinson and J. Wu, *Temperature-Adaptive Radiative Coating for All-Season Household Thermal Regulation*, Science **374**, 1504 (2021).

[2] M. D. Pickett, G. Medeiros-Ribeiro, and R. S. Williams, *A Scalable Neuristor Built with Mott Memristors*, Nat Mater **12**, 114 (2013).

[3] W. Yi, K. K. Tsang, S. K. Lam, X. Bai, J. A. Crowell, and E. A. Flores, *Biological Plausibility and Stochasticity in Scalable $VO_2$ Active Memristor Neurons*, Nat Commun **9**, 4661 (2018).

[4] P. Salev, J. Del Valle, Y. Kalcheim, and I. K. Schuller, *Giant Nonvolatile Resistive Switching in a Mott Oxide and Ferroelectric Hybrid*, Proc Natl Acad Sci USA **116**, 8798 (2019).

[5] A. Ronchi, P. Franceschini, P. Homm, M. Gandolfi, G. Ferrini, S. Pagliara, F. Banfi, M. Menghini, J. P. Locquet, and C. Giannetti, *Light-Assisted Resistance Collapse in a $V_2O_3$ -Based Mott-Insulator Device*, Phys. Rev. Appl. **15**, 044023 (2021).

[6] J. del Valle, N. Vargas, R. Rocco, P. Salev, Y. Kalcheim, P. N. Lapa, C. Adda, M.-H. Lee, P. Y. Wang, L. Fratino, M. J. Rozenberg, I. K. Schuller, *Spatiotemporal Characterization of the Field-Induced Insulator-to-Metal Transition*, Science **373**, 907 (2021).

[7] J. S. Brockman, L. Gao, B. Hughes, C. T. Rettner, M. G. Samant, K. P. Roche, and S. S. P. Parkin, *Subnanosecond Incubation Times for Electric-Field-Induced Metallization of a Correlated Electron Oxide*, Nat Nanotechnol **9**, 453 (2014).

[8] M. Lange, S. Guénon, Y. Kalcheim, T. Luibrand, N. M. Vargas, D. Schwebius, R. Kleiner, I. K. Schuller, and D. Koelle, *Imaging of Electrothermal Filament Formation in a Mott Insulator*, Phys Rev Appl **16**, 54027 (2021).

[9] S. Kumar, J. P. Strachan, and R. S. Williams, Chaotic Dynamics in Nanoscale $NbO_2$ Mott Memristors for Analogue Computing, Nature **548**, 318 (2017).

[10] J. del Valle, P. Salev, S. Gariglio, Y. Kalcheim, I. K. Schuller, and J.-M. Triscone, *Generation of Tunable Stochastic Sequences Using the Insulator-Metal Transition*, Nano Lett **22**, 1251 (2022).

[11] S. Cueff, J. John, Z. Zhang, J. Parra, J. Sun, R. Orobtchouk, S. Ramanathan, and P. Sanchis, *$VO_2$ Nanophotonics*, APL Photonics **5**, 110901 (2020).

[12] N. Manca, L. Pellegrino, T. Kanki, S. Yamasaki, H. Tanaka, A. S. Siri, and D. Marré, *Programmable Mechanical Resonances in MEMS by Localized Joule Heating of Phase Change Materials*, Advanced Materials **25**, 6430 (2013).

[13] D. Lee, B. Chung, Y. Shi, G.-Y. Kim, N. Campbell, F. Xue, K. Song, S.-Y. Choi, J. P. Podkaminer, T. H. Kim, P. J. Ryan, J.-W. Kim, T. R. Paudel, J.-H. Kang, J. W. Spinuzzi, D. A. Tenne, E. Y. Tsymbal, M. S. Rzchowski, L. Q. Chen, J. Lee and C. B. Eom, *Isostructural Metal-Insulator Transition in $VO_2$*, Science **362**, 1037 (2018).

[14] S. Biermann, A. Poteryaev, A. I. Lichtenstein, and A. Georges, Dynamical Singlets and Correlation-Assisted Peierls Transition in $VO_2$, Phys Rev Lett **94**, 026404 (2005).

[15] V. R. Morrison, R. P. Chatelain, K. L. Tiwari, A. Hendaoui, A. Bruhács, M. Chaker, and B. J. Siwick, *A Photoinduced Metal-like Phase of Monoclinic $VO_2$ Revealed by Ultrafast Electron Diffraction*, Science **346**, 445 (2014).

[16] L. Vidas, D. Schick, E. Martínez, D. Perez-Salinas, A. Ramos-Álvarez, S. Cichy, S. Batlle-Porro, A. S. Johnson, K. A. Hallman, R. F. Haglund, Jr. and S. Wall, *Does $VO_2$ Host a Transient Monoclinic Metallic Phase?* Phys Rev X **10**, 031047 (2020).

[17] S. Wall, S. Yang, L. Vidas, M. Chollet, J. M. Glownia, M. Kozina, T. Katayama, T. Henighan, M. Jiang, T. A. Miller, D. A. Reis, L. A. Boatner, O. Delaire and M. Trigo, *Ultrafast Disordering of Vanadium Dimers in Photoexcited $VO_2$*, Science **362**, 572 (2018).

[18] M. Thees, M.-H. Lee, R. L. Bouwmeester, P.H. Rezende-Gocalves, E. David, A. Zimmers, F. Fortuna, E. Frantzeskakis, N. M. Vargas, Y. Kalcheim, P. Le Fèvre, K. Horiba, H. Kumigashira, S. Biermann, J. Trastoy, M. J. Rozenberg, I. K. Schuller and A. F. Santander-Syro, *Imaging the Itinerant-to-Localized Transmutation of Electrons across the Metal-to-Insulator Transition in $V_2O_3$*, Sci Adv **7**, 1164 (2021).

[19] S. Lupi, L. Baldassarre, B. Mansart, A. Perucchi, A. Barinov, P. Dudin, E. Papalazarou, F. Rodolakis, J.-P. Rueff, J.-P. Itié, S. Ravy, D. Nicoletti, P. Postorino, P. Hansmann, N. Parragh, A. Toschi, T. Saha-Dasgupta, O. K. Andersen, G. Sangiovanni, K. Held and M. Marsi, *A Microscopic View on the Mott Transition in Chromium-Doped $V_2O_3$*, Nat Commun **1**, 105 (2010).

[20] A. S. McLeod, E. van Heumen, J. G. Ramirez, S. Wang, T. Saerbeck, S. Guenon, M. Goldflam, L. Anderegg, P. Kelly, A. Mueller, M. K. Liu, Ivan K. Schuller and D. N. Basov, *Nanotextured Phase Coexistence in the Correlated Insulator $V_2O_3$*, Nat Phys **13**, 80 (2017).

[21] Y. Kalcheim, N. Butakov, N. M. Vargas, M. H. Lee, J. Del Valle, J. Trastoy, P. Salev, J. Schuller, and I. K. Schuller, *Robust Coupling between Structural and Electronic Transitions in a Mott Material*, Phys Rev Lett **122**, 057601 (2019).

[22] Y. Ding, C.-C. Chen, Q. Zeng, H.-S. Kim, M. J. Han, M. Balasubramanian, R. Gordon, F. Li, L. Bai, D. Popov, S. M. Heald, T. Gog, H-K. Mao and M. van Veenendaal, *Novel High-Pressure Monoclinic Metallic Phase of $V_2O_3$*, Phys Rev Lett **112**, 056401 (2014).

[23] M. Imada, A. Fujimori, and Y. Tokura, *Metal-Insulator Transitions*, Rev Mod Phys **70**, 1039 (1998).

[24] F. J. Morin, *Oxides Which Show a Metal-to-Insulator Transition at the Neel Temperature*, Phys Rev Lett **3**, 34 (1959).

[25] D.B. McWhan and J. P. Remeika, *Metal-Insulator Transition in $V_{1-x}Cr_xO_3$*, Phys Rev B **2**, 3734 (1970).

[26] Y. Kalcheim, C. Adda, P. Salev, M. H. Lee, N. Ghazikhanian, N. M. Vargas, J. del Valle, and I. K. Schuller, *Structural Manipulation of Phase Transitions by Self-Induced Strain in Geometrically Confined Thin Films*, Adv Funct Mater **30**, 2005939 (2020).

[27] M. M. Qazilbash, M. Brehm, B.-G. Chae, P.-C. Ho, G. O. Andreev, B.-J. Kim, S. J. Yun, A. V. Balatsky, M. B. Maple, F. Keilmann, H.-T. Kim and D. N. Basov, *Mott Transition in $VO_2$ Revealed by Infrared Spectroscopy and Nano-Imaging*, Science **318**, 1750 (2007).

[28] G. Mattoni, P. Zubko, F. Maccherozzi, A.J.H. van der Torren, D. B. Boltje, M. Hadjimichael, N. Manca, S. Catalano, M. Gibert, Y. Liu, J. Aarts, J.-M. Triscone, S. S.

Dhesi and A. D. Caviglia, *Striped Nanoscale Phase Separation at the Metal-Insulator Transition of Heteroepitaxial Nickelates*, Nat Commun **7**, 13141(2016).

[29] J. H. Lee, F. Trier, T. Cornelissen, D. Preziosi, K. Bouzehouane, S. Fusil, S. Valencia, and M. Bibes, *Imaging and Harnessing Percolation at the Metal-Insulator Transition of NdNiO$_3$ Nanogaps*, Nano Lett **19**, 7801 (2019).

[30] L. Zhang, C. Israel, A. Biswas, R. L. Greene, and A. De Lozanne, *Direct Observation of Percolation in a Manganite Thin Film*, Science **298**, 805 (2002).

[31] C. Baldasseroni, C. Bordel, A. X. Gray, A. M. Kaiser, F. Kronast, J. Herrero-Albillos, C. M. Schneider, C. S. Fadley, and F. Hellman, *Temperature-Driven Nucleation of Ferromagnetic Domains in FeRh Thin Films*, Appl Phys Lett **100**, 262401(2012).

[32] E. Barazani, D. Das, C. Huang, A. Rakshit, C. Saguy, P. Salev, J. del Valle, M. C. Toroker, I. K. Schuller, and Y. Kalcheim, *Positive and Negative Pressure Regimes in Anisotropically Strained V$_2$O$_3$ Films*, Adv Funct Mater **33**, 2211801 (2023).

[33] J. Laverock, S. Kittiwatanakul, A. Zakharov, Y. Niu, B. Chen, S. A. Wolf, J. W. Lu, and K. E. Smith, *Direct Observation of Decoupled Structural and Electronic Transitions and an Ambient Pressure Monocliniclike Metallic Phase of VO$_2$*, Phys Rev Lett **113**, 216402 (2014).

[34] F. Mazzola, S. Kumar Chaluvadi, V. Polewczyk, D. Mondal, J. Fujii, P. Rajak, M. Islam, R. Ciancio, L. Barba, M. Fabrizio, G. Rossi, P. Orgiani and I. Vobornik, *Disentangling Structural and Electronic Properties in V$_2$O$_3$ Thin Films: A Genuine Nonsymmetry Breaking Mott Transition*, Nano Lett **22**, 5990 (2022).

[35] A. Ronchi, P. Franceschini, A. De Poli, P. Homm, A. Fitzpatrick, F. Maccherozzi, G. Ferrini, F. Banfi, S. S Dhesi, M. Menghini, M. Fabrizio, J.-P. Locquet and G. Giannetti, *Nanoscale Self-Organization and Metastable Non-Thermal Metallicity in Mott Insulators*, Nat Commun **13**, 3730 (2022).

[36] A. Ronchi, P. Homm, M. Menghini, P. Franceschini, F. Maccherozzi, F. Banfi, G. Ferrini, F. Cilento, F. Parmigiani, S. S. Dhesi, M. Fabrizio, J.-P. Locquet, and C. Giannetti, *Early-Stage Dynamics of Metallic Droplets Embedded in the Nanotextured Mott Insulating Phase of V$_2$O$_3$*, Phys Rev B **100**, 075111 (2019).

[37] S. V. Ovsyannikov, D. M. Trots, A. V. Kurnosov, W. Morgenroth, H. P. Liermann, and L. Dubrovinsky, *Anomalous Compression and New High-Pressure Phases of Vanadium Sesquioxide, V$_2$O$_3$*, J. Phys. Cond. Matt. **25**, 385401 (2013).

[38] M. Lucht, M. Lerche, H. C. Wille, Y. V. Shvyd'Ko, H. D. Rüter, E. Gerdau, and P. Becker, *Precise Measurement of the Lattice Parameters of α-Al$_2$O$_3$ in the Temperature Range 4.5-250 K Using the Mössbauer Wavelength Standard*, J Appl Crystallogr **36**, 1075 (2003).

[39] J.-H. Park, L. H. Tjeng, A. Tanaka, J. W. Allen, C. T. Chen, P. Metcalf, J. M. Honig, F. M. F. de Groot, and G. A. Sawatzky, Spin and Orbital Occupation and Phase Transitions in V$_2$O$_3$, Phys Rev B **61**, 11506 (2000).

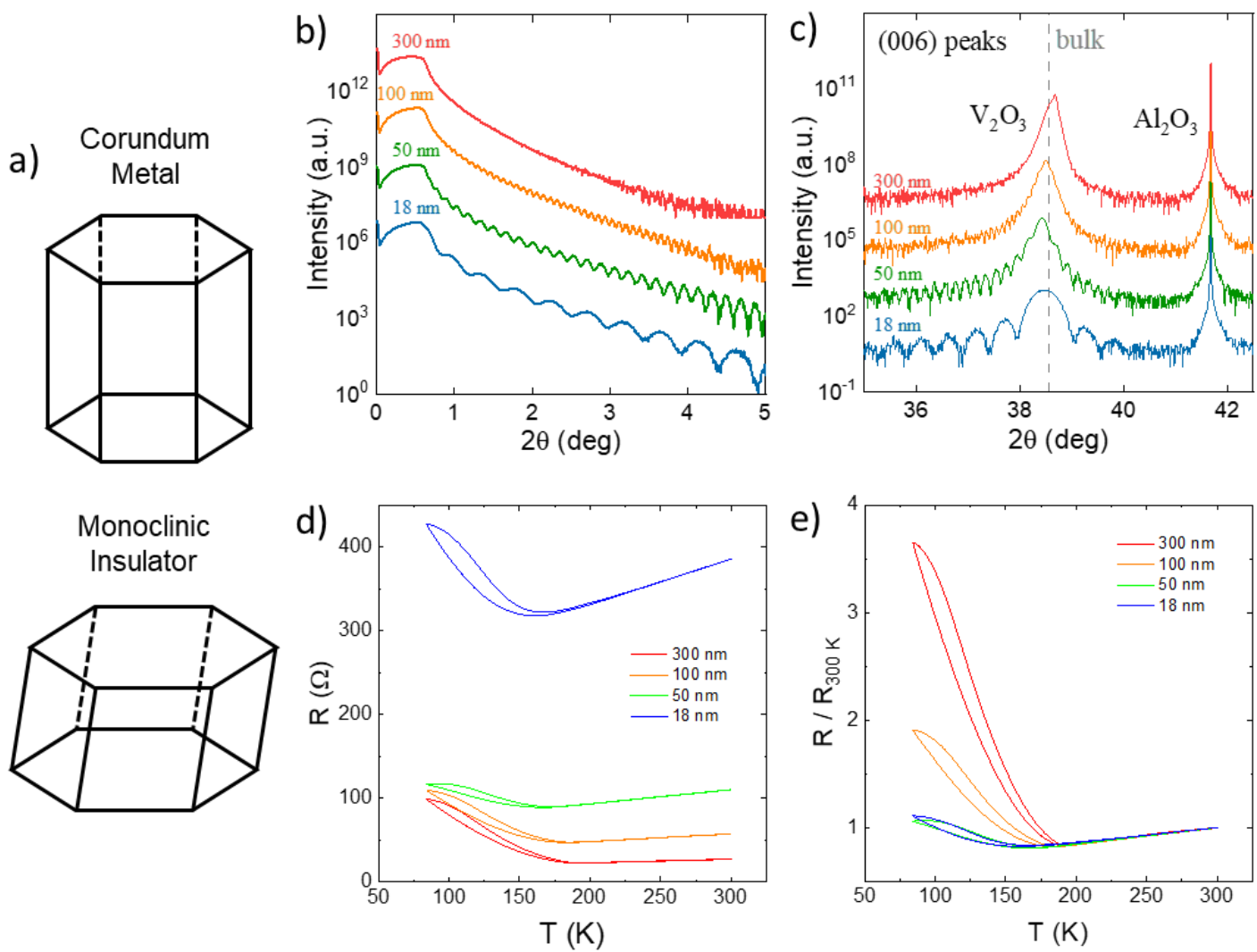

FIG. 1. (a) Schematic representation of the $V_2O_3$ unit cell in the corundum and monoclinic phases. (b) X-ray reflectivity of four (001)-oriented $V_2O_3$ films grown on (001)-oriented sapphire, with thicknesses ranging between 18 and 300 nm. (c) θ-2θ scans around the 006 peaks of the $V_2O_3$ films and the sapphire substrates. The vertical dashed line shows the position of the (006) peak for relaxed bulk $V_2O_3$. (d) Two-point resistance vs temperature for the four $V_2O_3$ films. (e) Resistance normalized by the film´s resistance at 300 K, to better compare the MIT suppression.

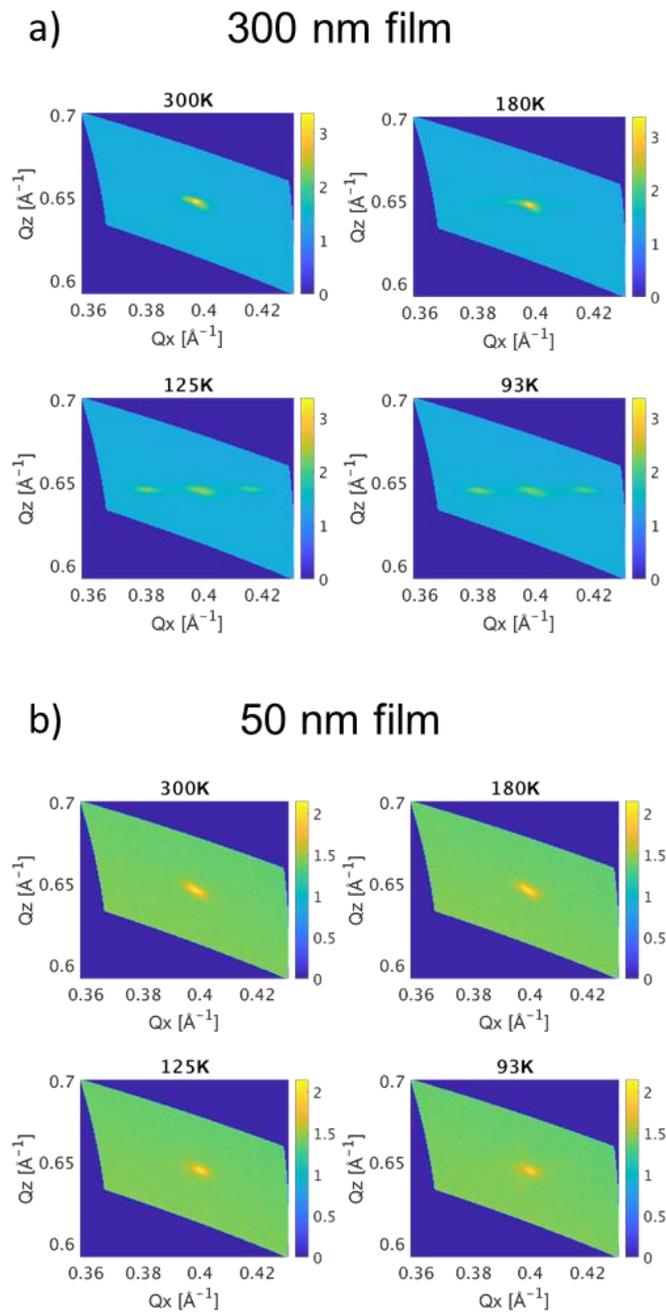

FIG. 2. Reciprocal space maps around the 119 direction for (a) a 300 nm film and (b) a 50 nm film. Several temperatures are shown.

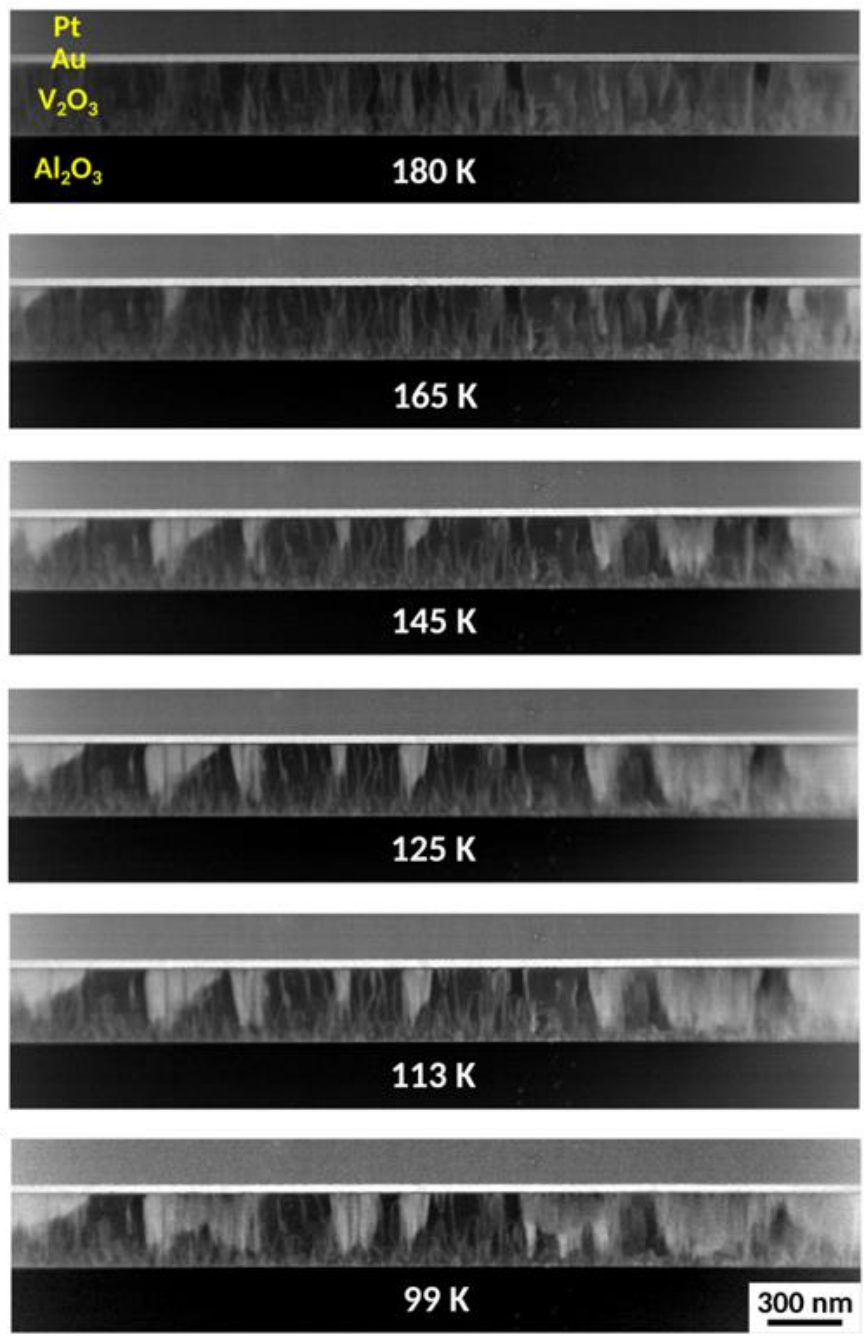

FIG. 3. Cryogenic STEM LAADF electron micrographs of the 300 nm thick $V_2O_3$ film from 180 K to 99 K. Because LAADF imaging in STEM is very sensitive to lattice distortion (largely diffraction contrast), it is used here to reveal crystalline defects, grain boundaries and thickness variation as well as to dissociate the corundum and the monoclinic crystal structure of $V_2O_3$. The extended bright and dark areas correspond to the monoclinic and the corundum phases of the $V_2O_3$ material, respectively.

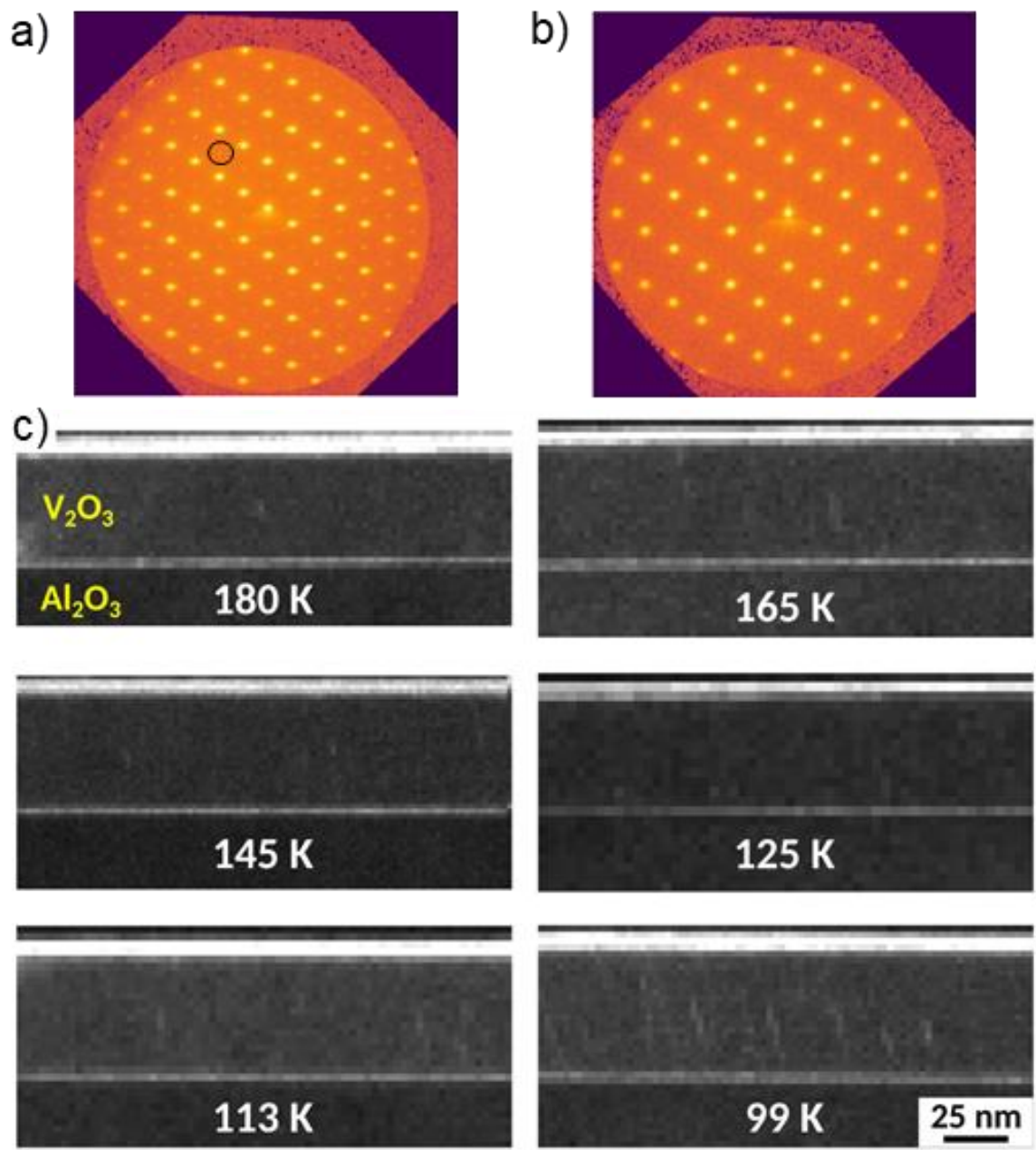

FIG. 4. 4D-STEM characterization of the $V_2O_3$ films at cryogenic temperature. a) Locally averaged electron diffraction pattern in log scale recorded at 125 K from a monoclinic $V_2O_3$ region in the 300 nm thick film (indexed using the corundum phase notation). b) Locally averaged electron diffraction pattern in log scale recorded at 125 K from the 50 nm thick $V_2O_3$ film. c) Intensity map from the circled areas in a) capturing the singularity from the monoclinic phase in the 50 nm $V_2O_3$ film, from 180 K to 99 K.

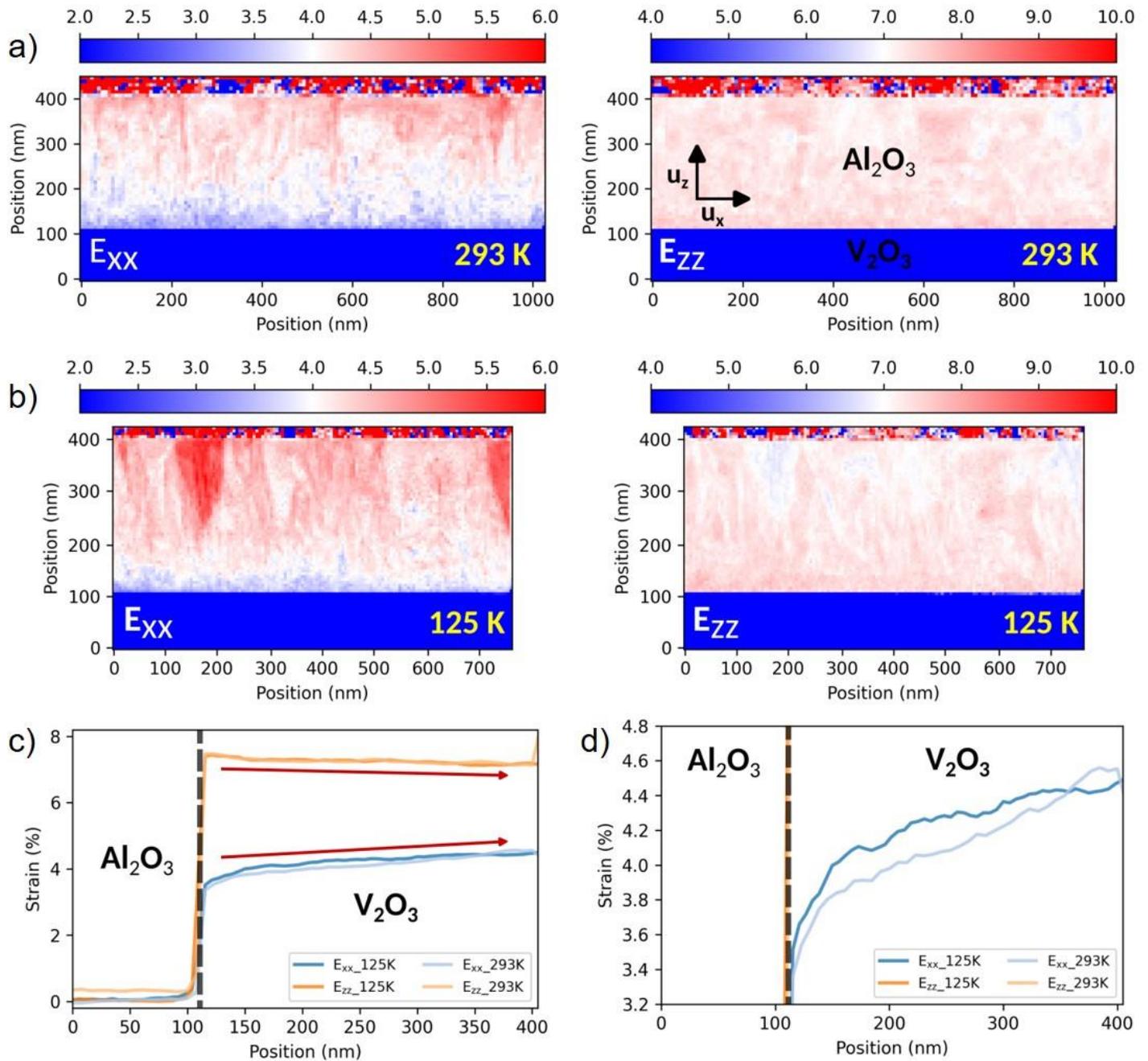

FIG. 5. Strain characterization of the 300 nm thick $V_2O_3$ film at (a) 293 K and (b) 125 K. The different panels correspond to uniaxial relative deformation maps ($E_{xx}$, $E_{zz}$) of the $V_2O_3$ film, using the sapphire substrate as the reference. (c) Vertical $E_{xx}$ and $E_{zz}$ profile lines, integrated along the x axis. Both 293 K and 125 K are shown. The black dashed line shows where the interface between substrate and film is. (d) Enlarged plot of the $E_{xx}$ components from (c). Compressive strain increases as the interface is approached.

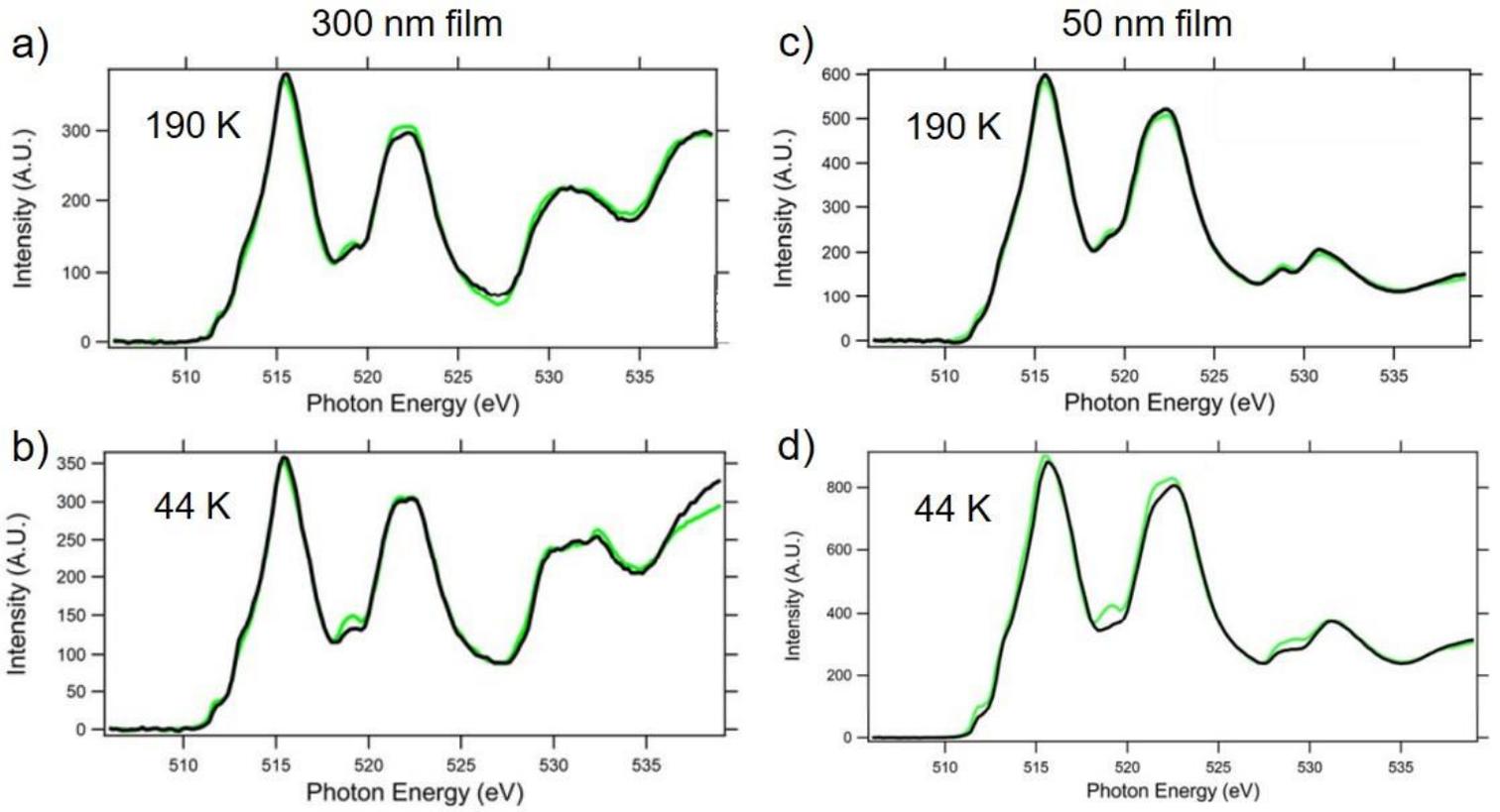

FIG. 6. X-Ray absorption near the vanadium L and Oxygen K edges for the (a) 300 nm film at 190 K, (b) 300 nm film at 44 K, (c) 50 nm film at 190 K and (d) 50 nm film at 44 K. For each panel, scans with light polarized in plane (black) and out of plane (green) are shown.

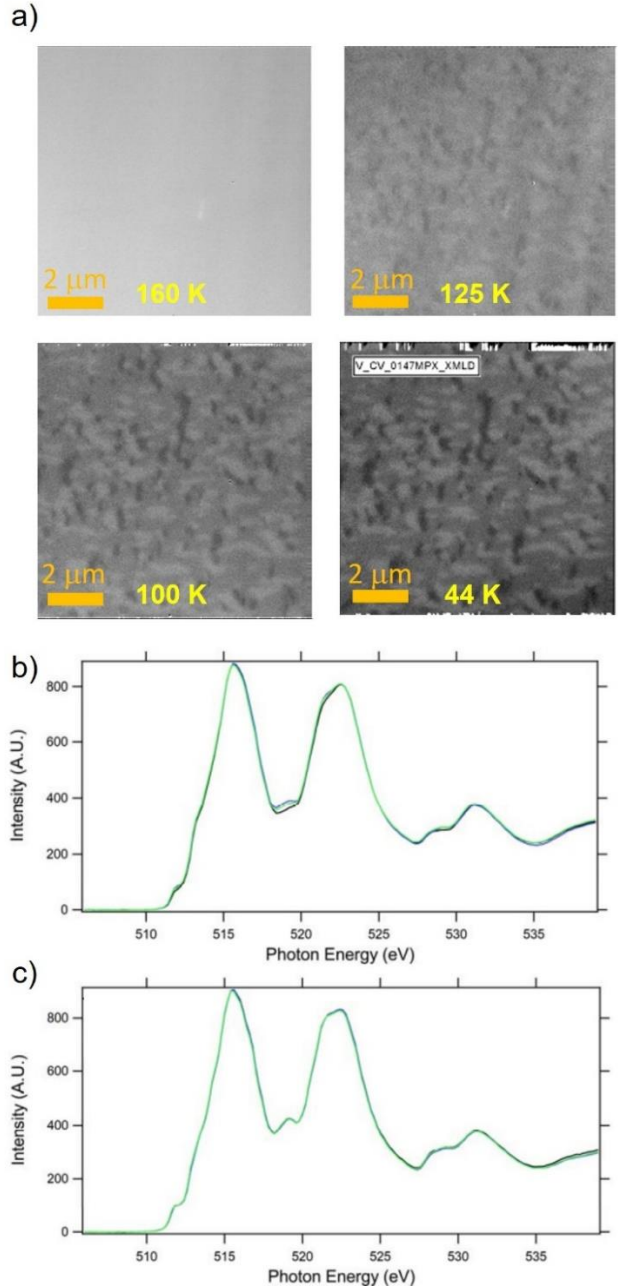

FIG. 7 (a) XLD (V/H) images of the 50 nm film with an incoming beam energy of 519.2 eV. Orange scale bars are 2 μm long. Several temperatures are shown. Temperature scanning was done by increasing temperature. Cool down measurements are shown in Fig. S4. (b) X-Ray absorption for the 50 nm film at 44 K with light polarized in plane. The three scans correspond to three regions of interest (contrast domains) in the PEEM map. These were chosen in accordance with the three domain intensities observed. (c) X-Ray absorption for the 50 nm film at the 44 K with light polarized out of plane. The three scans correspond to three regions of interest (contrast domains) observed in the PEEM map. These were chosen in accordance with the three intensity domain observed.

# Supplementary information

## 1. STEM measurements

Thin films were prepared into approximately 120 nm thick lamellas using the Helios 600 DualBeam instrument operating at 30 keV with a final clean-up at 5 keV. Samples were observed in a double aberration corrected JEOL ARM 200F cold FEG microscope operating at 200 keV in STEM mode with a semi-convergence angle of 21 mrad and a current of 60 pA resulting in a probe size of approximately 100 pm. The Double-tilt LN2 Atmos Defend Holder from Mel-Build was used for the cryogenic experiment from 300 K down to 99 K.

Figure 3 shows STEM Low Angle Annular Dark Field (LAADF) images of the 300 nm $V_2O_3$ film using a camera length of 20 cm corresponding to an inner and outer angle of 27 mrad and 110 mrad, respectively. This imaging mode favors the collection of the diffracted beams and is especially sensitive to defects, grain boundaries, thickness variations and structural phases. Since the intensity distribution of the diffracted peaks is different between the monoclinic and the corundum phase of the $V_2O_3$ material, the intensity collected in the ADF detector varies. As a result, the contrast in the STEM LAADF image directly distinguishes the corundum (dark) from the monoclinic (bright) phases in the $V_2O_3$ film. A weaker contrast variation is also visible on top of the structural one and corresponds to defects/domains resulting from the columnar vertical film growth observed at all temperatures.

To completely discard the presence of monoclinic regions within the bulk of the film, we used the 4D-STEM method, which allowed us to obtain a locally averaged diffraction pattern with a 5 nm spatial resolution and a convergence angle of around 0.85 mrad. The experimental setup was made using a custom alignment of the JEOL ARM optics and the condenser lens aperture of 5 um. The diffraction patterns were recorded on a Gatan K3 camera in counting mode and filtered using a 10-eV slit from the GIF Continuum spectrometer.

We used the 4D-STEM data to perform strain analysis using an adapted routine from the py4DSTEM v0.14.3 python package. (B. H. Savitzky *et al*. Microscopy and Microanalysis **27**, 712, 2021).

## 2. Supplementary figures

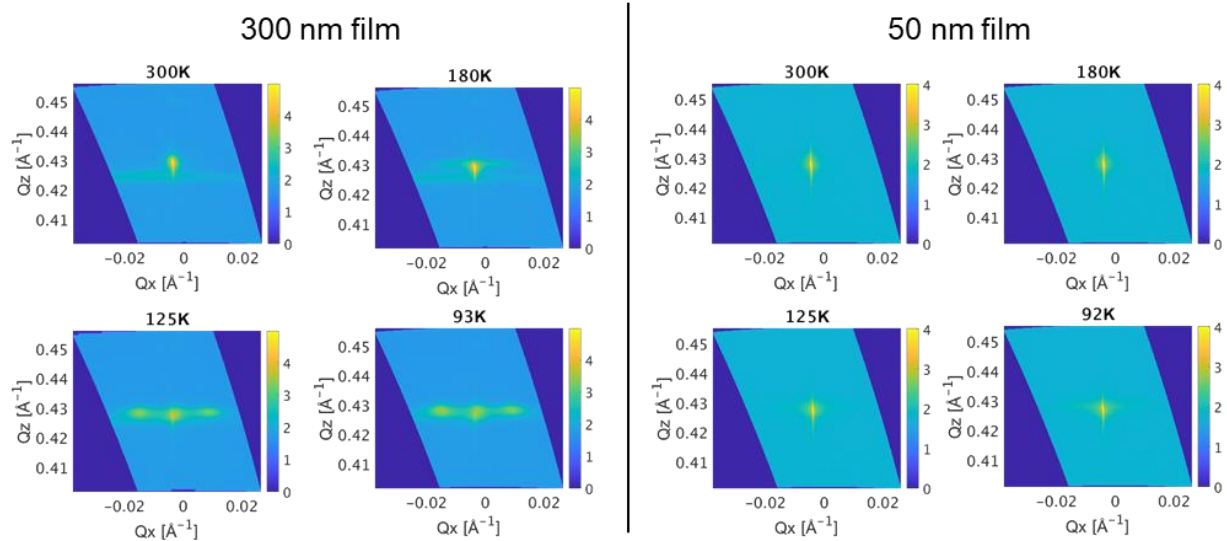

FIG. S1. Reciprocal space maps around the 006 direction for the 300 nm film and 50 nm films. Several temperatures are shown.

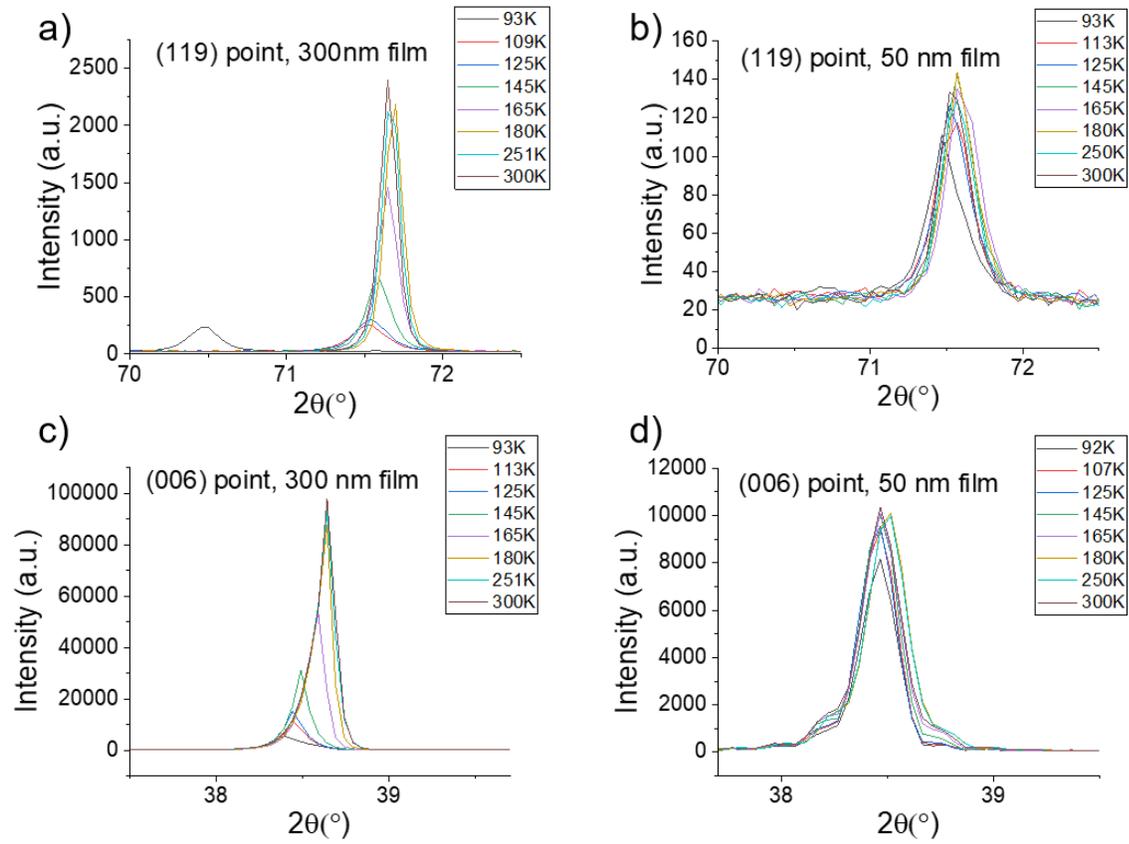

FIG. S2. θ-2θ scans around: (a) the 119 point for the 300 nm film, (b) the 119 point for the 50 nm film, (c) the 006 point for the 300 nm film, and (d) the 006 point for the 50 nm film. Several temperatures are shown in each plot.

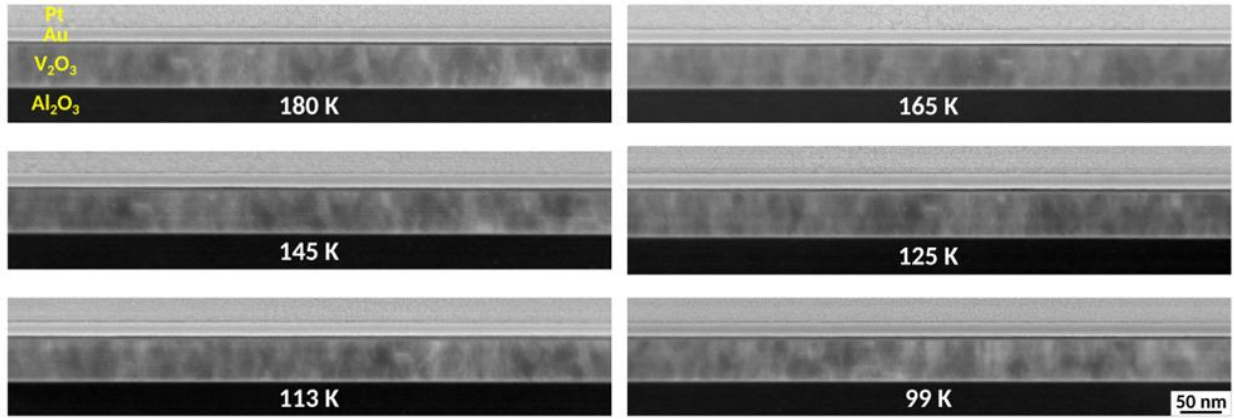

FIG. S3. Cryogenic STEM LAADF electron micrographs of the 50 nm thick $V_2O_3$ film from 180 K to 99 K. No monoclinic $V_2O_3$ was observed in the film.

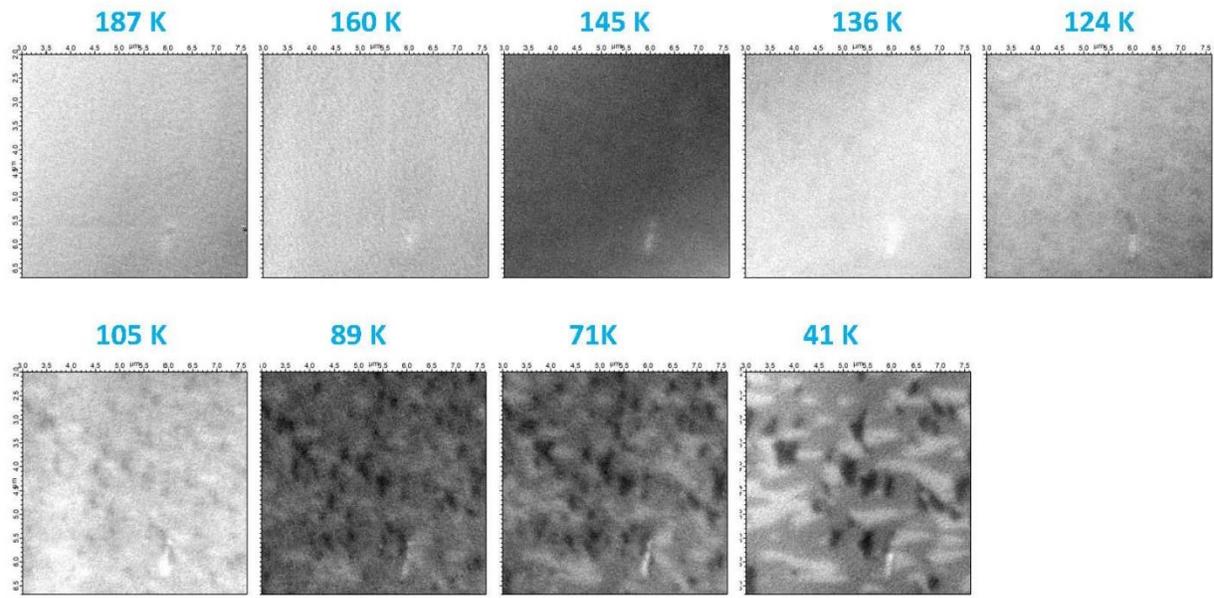

FIG. S4. XLD (V/H) images of the 50 nm film with an incoming beam energy of 519.2 eV. Several temperatures are shown. Temperature scanning was done by cooling down from 187 K to 41 K.

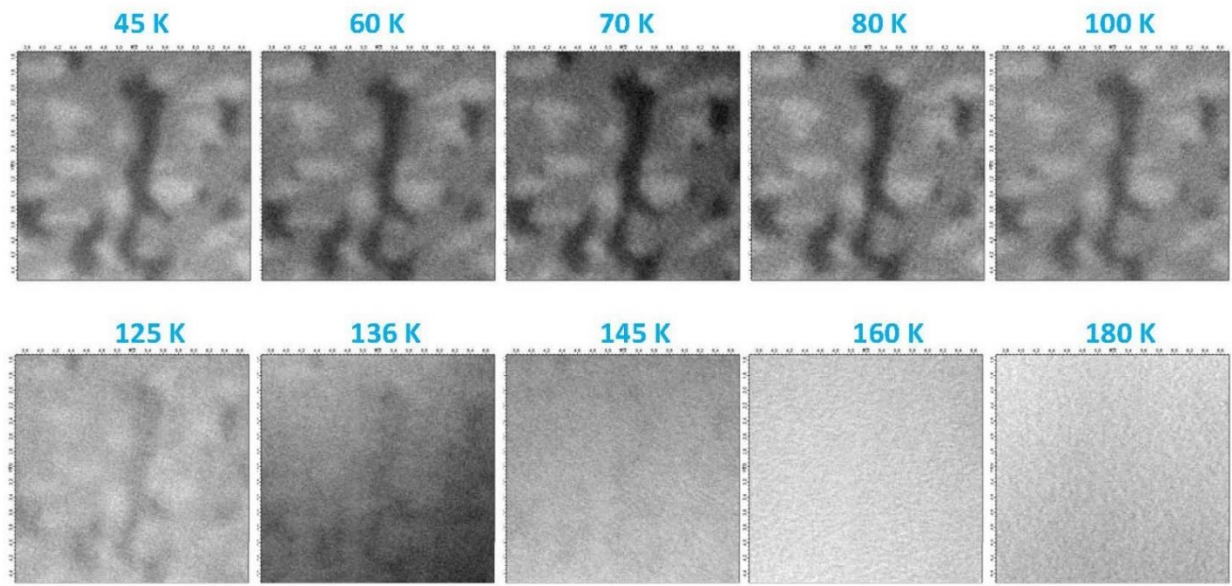

FIG. S5 XLD image zoom into a specific domain of the 50 nm film, showing little shrinkage during warm up.